\documentclass[10pt,letterpaper]{article}
\usepackage[top=0.85in,left=1.00in,footskip=0.75in]{geometry}
\usepackage[utf8]{inputenc}
\usepackage{amsmath,amssymb}

\usepackage{mathtools}

\usepackage{changepage}

\usepackage{color}
\usepackage[table,xcdraw,dvipsnames]{xcolor}
\definecolor{cuteBlue}{rgb}{0.258, 0.387, 0.574}
\definecolor{color1}{rgb}{0.0, 0.16, 0.67}
\definecolor{color2}{rgb}{0.3, 0.0, 0.3}
\definecolor{color3}{rgb}{0.0, 0.33, 0.0}

\usepackage{nameref}
\usepackage[colorlinks=true, urlcolor=cuteBlue, citecolor=black, linkcolor=black]{hyperref}

\usepackage{microtype}
\DisableLigatures[f]{encoding = *, family = *}

\usepackage[aboveskip=1pt,labelfont=bf,labelsep=period,justification=raggedright,singlelinecheck=off]{caption}

\usepackage{xparse}

\DeclareDocumentCommand \eref{oooo} {\IfNoValueTF{#2}{Eq.~\ref{#1}}{\IfNoValueTF{#3}{Eqs.~\ref{#1} and \ref{#2}}{\IfNoValueTF{#4}{Eqs.~\ref{#1}-\ref{#3}}{Eqs.~\ref{#1}-\ref{#4}}}}}

\DeclareDocumentCommand \tref{oooo} {\IfNoValueTF{#2}{Table~\ref{#1}}{\IfNoValueTF{#3}{Tables~\ref{#1} and \ref{#2}}{\IfNoValueTF{#4}{Tables~\ref{#1}-\ref{#3}}{Tables~\ref{#1}-\ref{#4}}}}}

\DeclareDocumentCommand \fref{ooo} {\IfNoValueTF{#2}{Fig.~\ref{#1}}{\IfNoValueTF{#3}{Figs.~\ref{#1} and \ref{#2}}{Figs.~\ref{#1}-\ref{#3}}}}

\newcommand{\letter}[1]{#1} 
\newcommand{\letterParen}[1]{(#1)} 

\usepackage{graphicx}
\usepackage{multirow}

\usepackage{units, xfrac}

\mathchardef\mhyphen="2D

\usepackage{accents}

\newif\ifShowALL
\newif\ifShowSI

\usepackage[
backend=bibtex,
citestyle=numeric-comp, 
sorting=none, 			
url=false, 				
doi=false, 				
isbn=false, 			
eprint=false, 			
maxcitenames=99, 		
maxbibnames=99,			
firstinits=true
]{biblatex}
\DeclareFieldFormat{labelnumberwidth}{#1\adddot}

\renewbibmacro{in:}{%
	\ifentrytype{article}{}{%
		\printtext{\bibstring{in}\intitlepunct}%
	}%
}	

\renewbibmacro*{volume+number+eid}{
	\printfield{volume}
	\printunit{\addcolon}
}

\renewbibmacro*{journal+issuetitle}{%
	\usebibmacro{journal}%
	\setunit{\adddot\space}%
	\usebibmacro{volume+number+eid}%
}

\DeclareNameAlias{default}{last-first/first-last}

\renewbibmacro*{author}{%
	\printnames[][-\value{listtotal}]{author}%
	\setunit{\adddot\space}\newblock%
}

\DeclareFieldFormat[article]{title}{#1}

\DeclareFieldFormat{date}{#1\setunit{\adddot\space}}

\DeclareFieldFormat{pages}{#1}

\DeclareBibliographyDriver{article}{%
	\usebibmacro{author}%
	\usebibmacro{date}%
	\usebibmacro{title}%
	\usebibmacro{journal}%
	\usebibmacro{volume+number+eid}%
	\usebibmacro{note+pages}%
	\usebibmacro{finentry}%
}

\DeclareFieldFormat[misc]{title}{#1}

\DeclareFieldFormat{url}{\url{#1}\adddot}

\renewbibmacro*{url+urldate}{%
	\ifentrytype{misc}{\usebibmacro{url}}{}%
} 

\DeclareBibliographyDriver{misc}{%
	\usebibmacro{author}%
	\usebibmacro{date}%
	\usebibmacro{title}%
	\usebibmacro{url+urldate}%
	\usebibmacro{finentry}%
}

\begin{document}
	
\begin{flushleft} 
	{\Large \textbf\newline{How the Avidity of Polymerase Binding to the -35/-10 Promoter Sites Affects Gene Expression}}
	\newline
	\\
	\textbf{Tal Einav$^{1,*}$, Rob Phillips$^{1,2,3,*}$}
		\\
		$^1$Department of Physics, California Institute of Technology,
		Pasadena, CA, 91125, USA\\
		$^2$Department of Applied Physics, California Institute of Technology,
		Pasadena, CA, 91125, USA\\
		$^3$Division of Biology and Biological Engineering, California Institute of Technology,
		Pasadena, CA, 91125, USA\\
		*Corresponding author: tal.einav@caltech.edu; (512) 468-1994; 1200 E. California Blvd, MC 103-33, Pasadena, CA 91125
		*Corresponding author: phillips@pboc.caltech.edu; (626) 395-3374; 1200 E. California Blvd, MC 128-95, Pasadena, CA 91125
		
\end{flushleft}

\section*{Abstract}

Although the key promoter elements necessary to drive transcription in
\textit{Escherichia coli} have long been understood, we still cannot predict the
behavior of arbitrary novel promoters, hampering our ability to characterize the
myriad of sequenced regulatory architectures as well as to design new
synthetic circuits. This work builds on a beautiful recent experiment by Urtecho
\textit{et al.}~who measured the gene expression of over 10,000 promoters
spanning all possible combinations of a small set of regulatory elements. Using
this data, we demonstrate that a central claim in energy matrix models of gene
expression -- that each promoter element contributes independently and
additively to gene expression -- contradicts experimental measurements. We
propose that a key missing ingredient from such models is the avidity between
the -35 and -10 RNA polymerase binding sites and develop what we call a
\textit{refined energy matrix} model that incorporates this effect. We show that
this the refined energy matrix model can characterize the full suite of gene
expression data and explore several applications of this framework, namely, how
multivalent binding at the -35 and -10 sites can buffer RNAP kinetics against
mutations and how promoters that bind overly tightly to RNA polymerase can
inhibit gene expression. The success of our approach suggests that avidity
represents a key physical principle governing the interaction of RNA polymerase
to its promoter.

\section*{Significance Statement}

Cellular behavior is ultimately governed by the genetic program encoded in its
DNA and through the arsenal of molecular machines that actively transcribe its
genes, yet we lack the ability to predict how an arbitrary DNA sequence will
perform. To that end, we analyze the performance of over 10,000 regulatory
sequences and develop a model that can predict the behavior of any sequence
based on its composition. By considering promoters that only vary by one or two
elements, we can characterize how different components interact, providing
fundamental insights into the mechanisms of transcription.

\pagebreak
\section*{Introduction}

Promoters modulate the complex interplay of RNA polymerase (RNAP) and
transcription factor binding that ultimately regulates gene expression. While
our knowledge of the molecular players that mediate these processes constantly
improves, more than half of all promoters in \textit{Escherichia coli} still have no
annotated transcription factors in RegulonDB \cite{Gama-Castro2016} and our
ability to design novel promoters that elicit a target level of gene expression
remains limited.

As a step towards taming the vastness and complexity of sequence space, the
recent development of massively parallel reporter assays has enabled entire
libraries of promoter mutants to be simultaneously measured
\cite{Patwardhan2009, Kinney2010, Inoue2015}. Given this surge in experimental
prowess, the time is ripe to reexamine how well our models of gene expression
can extrapolate the response of a general promoter.

A common approach to quantifying gene expression, called the \textit{energy
	matrix model}, assumes that every promoter element contributes additively and
independently to the total RNAP (or transcription factor) binding energy
\cite{Kinney2010}. This model treats all base pairs on an equal footing and does
not incorporate mechanistic details of RNAP-promoter interactions such as its
strong binding primarily at the -35 and -10 binding motifs (shown in
\fref[figRNAPoverview]\letter{A}). A newer method recently took the opposite
viewpoint, designing an RNAP energy matrix that only includes the -35 element,
-10 element, and the length of the spacer separating them \cite{Yona2018},
neglecting the sequence composition of the spacer or the surrounding promoter
region.

Although these methods have been successfully used to identify important
regulatory elements in unannotated promoters \cite{Belliveau2018} and predict
evolutionary trajectories \cite{Yona2018}, it is clear that there is more to the
story. Even in the simple case of the highly-studied \textit{lac} promoter,
such energy matrices show systematic deviations from measured levels of gene
expressions, indicating that some fundamental component of transcriptional
regulation is still missing \cite{Forcier2018}.

We propose that one failure of current models lies in their tacit assumption
that every promoter element contributes independently to the RNAP binding
energy. By naturally relaxing this assumption to include the important effects
of avidity, we can push beyond the traditional energy matrix analysis in several
key ways including: (\textit{i}) We can identify which promoter elements
contribute independently or cooperatively without recourse to fitting, thereby
building an unbiased mechanistic model for systems that bind at multiple sites.
(\textit{ii}) Applying this approach to RNAP-promoter binding reveals that the
-35 and -10 motifs bind cooperatively, a feature that we attribute to avidity.
Moreover, we show that models that instead assume the -35 and -10 elements
contribute additively and independently sharply contradict the available data.
(\textit{iii}) We show that the remaining promoter elements (the spacer, UP, and
background shown in \fref[figRNAPoverview]\letter{A}) do contribute
independently and additively to the RNAP binding energy and formulate the
corresponding model for transcriptional regulation that we call a
\textit{refined energy matrix model}. (\textit{iv}) We use this model to explore
how the interactions between the -35 and -10 elements can buffer RNAP kinetics
against mutations. (\textit{v}) We analyze a surprising feature of the data
where overly-tight RNAP-promoter binding can lead to decreased gene expression.
(\textit{vi}) We validate our model by analyzing the gene expression of over
10,000 promoters in \textit{E.~coli} recently published by Urtecho \textit{et
	al.}~\cite{Urtecho2018} and demonstrate that our framework markedly improves
upon the traditional energy matrix analysis.

While this work focuses on RNAP-promoter binding, its implications extend to general
regulatory architectures involving multiple tight-binding elements including
transcriptional activators that make contact with RNAP (CRP in the \textit{lac}
promoter \cite{Kuhlman2007}), transcription factors that oligomerize (as
recently identified for the \textit{xylE} promoter \cite{Belliveau2018}), and
transcription factors that bind to multiple sites on the promoter (DNA looping
mediated by the Lac repressor \cite{Boedicker2013a}). More generally, this
approach of categorizing which binding elements behave independently (without
resorting to fitting) can be applied to multivalent interactions in other
biological contexts including novel materials, scaffolds, and synthetic switches
\cite{Varner2015, Yan2018}.

\section*{Results}

\subsection*{The -35 and -10 Binding Sites give rise to Gene Expression that Defies Characterization as Independent and Additive Components}

Decades of research have shed light upon the exquisite biomolecular details
involved in bacterial transcriptional regulation via the family of RNAP $\sigma$
factors \cite{Feklistov2014}. In this work, we restrict our attention to the
$\sigma^{70}$ holoenzyme \cite{Urtecho2018}, the most active form under standard
\textit{E. coli} growth condition, whose interaction with a promoter includes
direct contact with the -35 and -10 motifs (two hexamers centered roughly 10 and
35 bases upstream of the transcription start site), a spacer region separating
these two motifs, an UP element just upstream of the -35 motif that anchors the
C-terminal domain ($\alpha$CTD) of RNAP, and the background promoter sequence
surrounding these elements.

Urtecho \textit{et al.}~constructed a library of promoters composed of every
combination of eight -35 motifs, eight -10 motifs, eight spacers, eight
backgrounds (BG), and three UP elements (\fref[figRNAPoverview]\letter{A})
\cite{Urtecho2018}. Each sequence was integrated at the same locus within the
\textit{E. coli} genome and transcription was quantified via DNA barcoding and
RNA sequencing. One of the three UP elements considered was the absence of an UP
binding motif, and this case will serve as the starting point for our analysis.

The traditional energy matrix approach used by Urtecho \textit{et al.}~posits
that every base pair of the promoter will contribute additively and
independently to the RNAP binding energy \cite{Urtecho2018}, which by
appropriately grouping base pairs is equivalent to stating that the free energy
of RNAP binding will be the sum of its contributions from the background,
spacer, -35, and -10 elements (see
Appendix A). Hence, the gene expression (GE) is given by the Boltzmann factor
\begin{equation} \label{eqGeneExpressionNoUpApproximate}
\text{GE} \propto e^{-\beta (E_\text{BG} + E_\text{Spacer} + E_{\mhyphen 35} + E_{\mhyphen 10})}.
\end{equation}
Note that all $E_j$s represent free energies (with an energetic and entropic
component); to see the explicit dependence on RNAP copy number, refer to
Appendix A. Fitting the 32 free energies (one for each background, spacer, -35,
and -10 element) and the constant of proportionality in
\eref[eqGeneExpressionNoUpApproximate] enables us to predict the expression of
$8 \times 8 \times 8 \times 8 = 4{,}096$ promoters.

\fref[figLogLinearModel]\letter{A} demonstrates that
\eref[eqGeneExpressionNoUpApproximate] leads to a poor characterization of these
promoters ($R^2 = 0.57$, parameter values listed in Appendix B), suggesting that
critical features of gene expression are missing from this model. One possible
resolution is to assumes that the level of gene expression saturates for very
strong promoters at $r_{\text{max}}$ and for very weak promoters at $r_0$
(caused by background noise or serendipitous near-consensus sequences
\cite{Yona2018}), namely,
\begin{equation} \label{eqGeneExpressionWithDenominator}
\text{GE} = \frac{r_0 + r_{\text{max}} e^{-\beta (E_\text{BG} + E_\text{Spacer} + E_{\mhyphen 35} + E_{\mhyphen 10})}}{1 + e^{-\beta (E_\text{BG} + E_\text{Spacer} + E_{\mhyphen 35} + E_{\mhyphen 10})}}.
\end{equation}
Since \eref[eqGeneExpressionWithDenominator] still assumes that each promoter
element contributes additively and independently to the total RNAP binding
energy, it also makes sharp predictions that markedly disagrees with the data
(see Appendix C). Inspired by these inconsistencies, we postulated that certain
promoter elements, most likely the -35 and -10 sites, may not contribute
synergistically to RNAP binding.

To that end, we consider a model for gene expression shown in
\fref[figRNAPoverview]\letter{B} where RNAP can separately bind to the -35 and
-10 sites. RNAP is assumed to elicit a large level of gene expression
$r_{\text{max}}$ when fully bound but the smaller level $r_0$ when unbound or
partially bound. Importantly, the Boltzmann weight of the fully bound state
contains the free energy $E_\text{int}$ representing the avidity of RNAP binding
to the -35 and -10 sites. Physically, avidity arises because unbound RNAP
binding to either the -35 or -10 sites gains energy but loses entropy, while
this singly bound RNAP attaching at the other (-10 or -35) site again gains
energy but loses much less entropy, as it was tethered in place rather than
floating in solution. Hence we expect $e^{-\beta E_\text{int}} \gg 1$, and
including this avidity term implies that RNAP no longer binds independently to
the -35 and -10 sites.

Our coarse-grained model of gene expression neglects the kinetic details of
transcription whereby RNAP transitions from the closed to open complex before
initiating transcription. Instead, we assume that there is a separation of
timescales between the fast process of RNAP binding/unbinding to the promoter
and the other processes that constitute transcription. In the
quasi-equilibrium framework shown in \fref[figRNAPoverview]\letter{B}, gene
expression is given by the average occupancy of RNAP in each of its states, namely,
\begin{equation} \label{eqGeneExpressionNoUp}
\text{GE} = \frac{r_0 + e^{-\beta (E_\text{BG} + E_\text{Spacer})} \left( r_0 e^{-\beta E_{\mhyphen 35}} + r_0 e^{-\beta E_{\mhyphen 10}} + r_{\text{max}} e^{-\beta \left( E_{\mhyphen 35} + E_{\mhyphen 10} + E_\text{int} \right)} \right)}{1 + e^{-\beta (E_\text{BG} + E_\text{Spacer})} \left( e^{-\beta E_{\mhyphen 35}} + e^{-\beta E_{\mhyphen 10}} + e^{-\beta (E_{\mhyphen 35} + E_{\mhyphen 10} + E_\text{int})} \right)}.
\end{equation}
We call this expression a refined energy matrix model since it reduces to the
energy matrix \eref[eqGeneExpressionNoUpApproximate] (with constant of
proportionality $r_{\text{max}} E^{-\beta E_\text{int}}$) in the limit where
gene expression is negligible when the RNAP is not bound ($r_0 \approx 0$) and
the promoter is sufficiently weak or the RNAP concentration is sufficiently
small that polymerase is most often in the unbound state (so that the
denominator $\approx 1$). The background and spacer are assumed to contribute to
RNAP binding in both the partially and fully bound states, an assumption that we
rigorously justify in Appendix D.

\begin{figure}[t]
	\centering \includegraphics{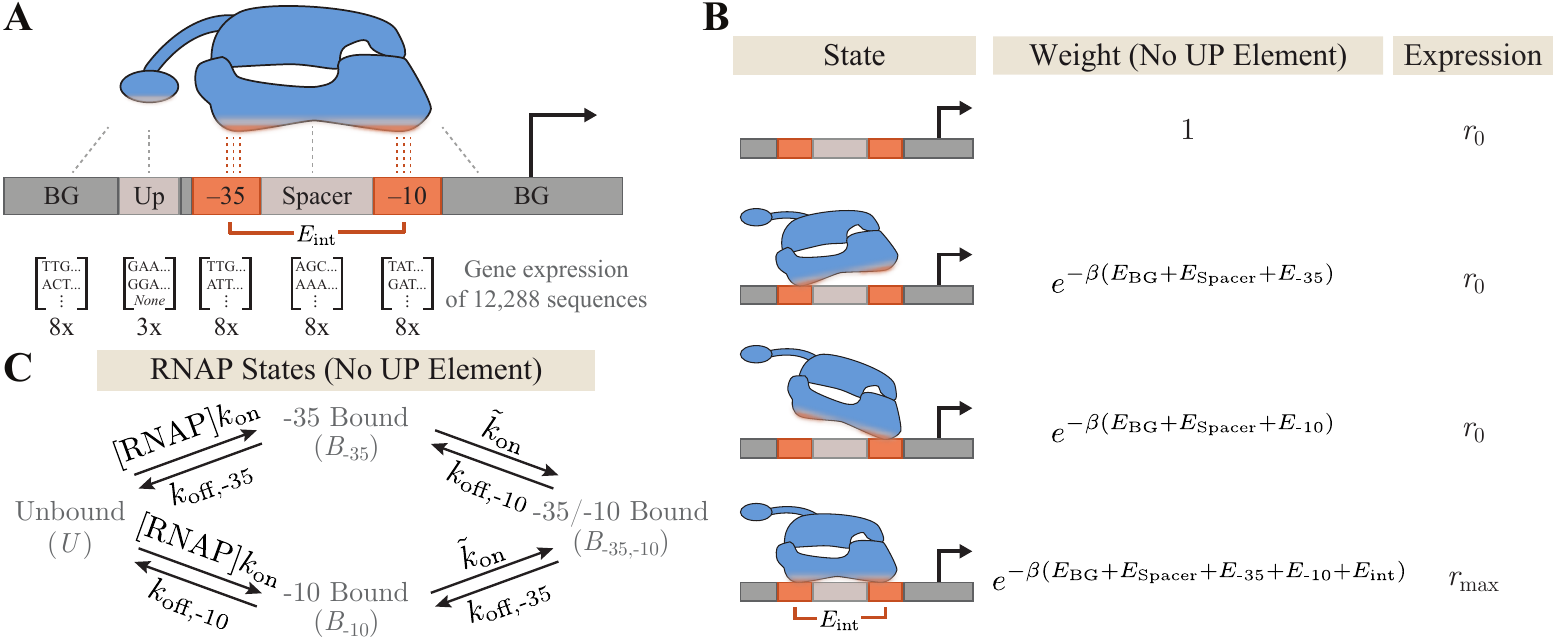}
	
	\caption{\textbf{The bivalent nature of RNAP-promoter binding.} \letterParen{A}
		Gene expression was measured for RNAP promoters comprising any combination of
		-35, -10, spacer, UP, and background (BG) elements. \letterParen{B} When no UP
		element is present, RNAP makes contact with the promoter at the -35 and -10
		sites giving rise to gene expression $r_0$ when unbound or partially bound and
		$r_{\text{max}}$ when fully bound. \letterParen{C} Having two binding sites alters the dynamics of
		RNAP binding. $k_{\text{on}}$ represents the on-rate from unbound to partially
		bound RNAP and $\tilde{k}_{\text{on}}$ the analogous rate from partially to
		fully bound RNAP, while $k_{\text{off}, j}$ denotes the unbinding rate from
		site $j$.} \label{figRNAPoverview}
\end{figure}

\begin{figure}[t!]
	\centering \includegraphics{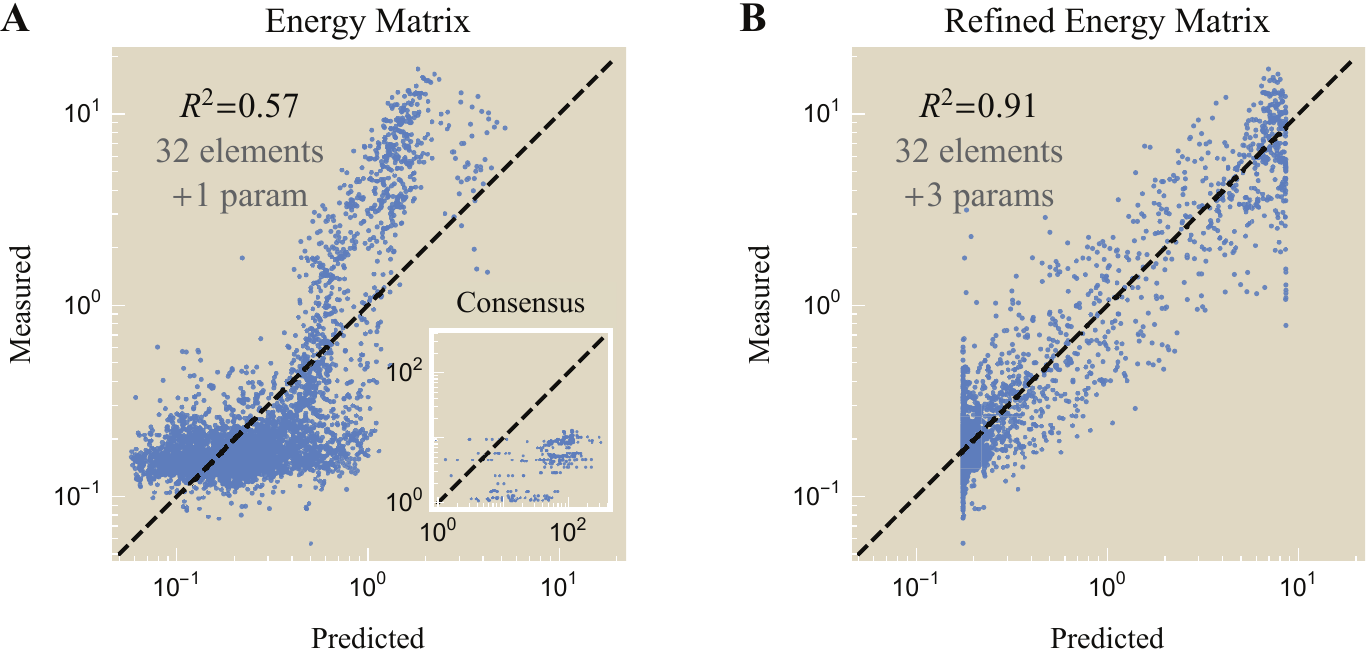}
	
	\caption{\textbf{Gene expression of promoters with no UP element.} Model
		predictions using \letterParen{A} an energy matrix
		(\eref[eqGeneExpressionNoUpApproximate]) where the -35 and -10 elements
		independently contribute to RNAP binding and \letterParen{B} a refined energy
		matrix (\eref[eqGeneExpressionNoUp]) where the two sites contribute
		cooperatively. Inset: The epistasis-free nature of the energy matrix model
		makes sharp predictions about the gene expression of the consensus -35 and -10
		sequences that markedly disagree with the data. Parameter values given in
		Appendix B.} \label{figLogLinearModel}
\end{figure}

\fref[figLogLinearModel]\letter{B} demonstrates that the refined energy matrix
\eref[eqGeneExpressionNoUp] is better able to capture the system's behavior
($R^2 = 0.91$) while only requiring two more parameters ($r_0$ and
$E_\text{int}$) than the energy matrix model
\eref[eqGeneExpressionNoUpApproximate]. The sharp boundaries on the left and
right represent the minimum and maximum levels of gene expression, $r_0 = 0.18$
and $r_\text{max} = 8.6$, respectively (see Appendix E). The refined energy
matrix predicts that the top 5\% of promoters will exhibit expression levels of
7.6 (compared to 8.5 measured experimentally) while the weakest 5\% of promoters
should express at 0.2 (compared to the experimentally measured 0.1). In
addition, this model quickly gains predictive power, as its coefficient of
determination only slightly diminishes ($R^2 = 0.86$) if the model is trained on
only 10\% of the data and used to predict the remaining 90\%.

\subsection*{Epistasis-Free Models of Gene Expression Lead to Sharp Predictions that Disagree with the Data}

To further validate that the lower coefficient of determination of the energy
matrix approach (\eref[eqGeneExpressionNoUpApproximate]) was not an artifact of
the fitting, we can utilize the epistasis-free nature of this model to predict
the gene expression of double mutants from that of single mutants. More
precisely, denote the gene expression $\text{GE}^{(0,0)}$ of a promoter with the
consensus -35 and -10 sequences (and any background or spacer sequence). Let
$\text{GE}^{(1,0)}$, $\text{GE}^{(0,1)}$, and $\text{GE}^{(1,1)}$ represent
promoters (with this same background and spacer) whose -35/-10 sequences are
mutated/consensus, consensus/mutated, and mutated/mutated, respectively, where
``mutated'' stands for any non-consensus sequence. As derived in Appendix D, the
gene expression of these three later sequences can predict the gene expression
of the promoter with the consensus -35 and -10 without recourse to fitting,
namely,
\begin{equation} \label{eqM10M35Interaction}
\text{GE}^{(0,0)} = \text{GE}^{(1,1)} \frac{\text{GE}^{(0,1)}}{\text{GE}^{(1,1)}} \frac{\text{GE}^{(1,0)}}{\text{GE}^{(1,1)}}.
\end{equation}
The inset in \fref[figLogLinearModel]\letter{A} compares the epistasis-free
predictions ($x$-axis, right-hand side of \eref[eqM10M35Interaction]) with the
measured gene expression ($y$-axis, left-hand side of
\eref[eqM10M35Interaction]). These results demonstrate that the simple energy
matrix formulation fails to capture the interaction between the -35 and -10
binding sites. While this calculation cannot readily generalize to the refined
energy matrix model since it exhibits epistasis, it is analytically tractable
for weak promoters where the refined energy matrix model displays a marked
improvement over the traditional energy matrix model (see Appendix C).

\subsection*{RNAP Binding to the UP Element occurs Independently of the Other Promoter Elements}

Having seen that the refined energy matrix model (\eref[eqGeneExpressionNoUp])
can outperform the traditional energy matrix analysis on promoters with no UP
element, we next extend the former model to promoters containing an UP element.
Given the importance of the RNAP interactions with the -35 and -10 sites seen
above, \fref[figThreeSiteModel]\letter{A} shows three possible mechanisms for
how the UP element could mediate RNAP binding. For example, the C-terminal could
bind strongly and independently so that RNAP has three distinct binding sites.
Another possibility is that the RNAP $\alpha$CTD binds if and only if the -35
binding site is bound. A third alternative is that the UP element contributes
additively and independently to RNAP binding (analogous to the spacer and
background).

To distinguish between these possibilities, we analyze the correlations in gene
expression between every pair of promoter elements (UP and -35, spacer and
background, etc.) to determine the strength of their interaction. Each model in
\fref[figThreeSiteModel]\letter{A} will have a different signature: The top
schematic predicts strong interactions between the -35 and -10, between the UP
and -35, and between the UP and -10; the middle schematic would give rise to
strong dependence between the -35 and -10 as well as between the UP and -10,
while the UP and -35 elements would be perfectly correlated; the bottom
schematic suggests that the UP elements will contribute independently of the
other promoter elements.

\begin{figure}[t!]
	\centering \includegraphics{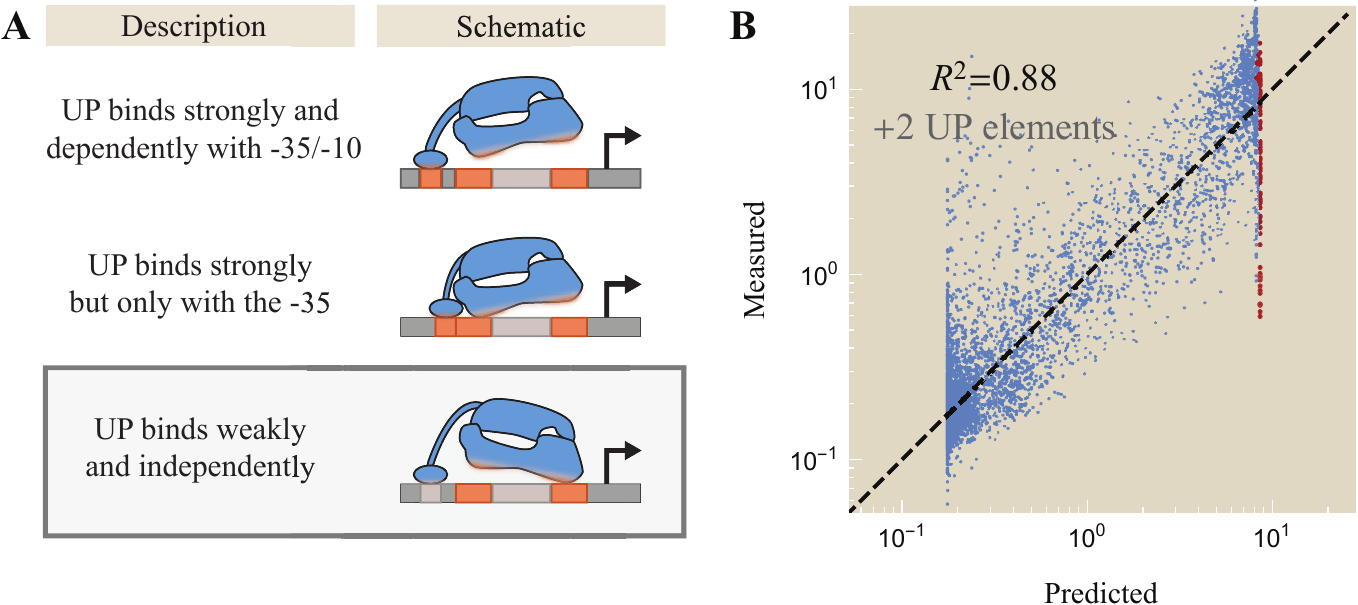}
	
		\caption{\textbf{The interaction between RNAP and the UP element.}
		\letterParen{A} Possible mechanisms by which the RNAP C-terminal can bind to
		the UP element (orange segments represent strong binding comparable to the -35
		and -10 motifs; gray segments represent weak binding comparable to the spacer
		and background). The data supports the bottom schematic (see Appendix D).
		\letterParen{B} The corresponding characterization of 8,192 promoters identical
		to those shown in \fref[figLogLinearModel] but with one of two UP binding
		motifs. Red points represent promoters with a consensus -35 and -10. Data was
		fit using the same parameters as in \fref[figLogLinearModel]\letter{B} and
		fitting the binding energies of the two UP elements (parameter values in
		Appendix B).} \label{figThreeSiteModel}
\end{figure}

This analysis, which we relegate to Appendix D, demonstrates that the UP element
is approximately independent of all other promoter elements ($R^2 \gtrsim 0.6$)
as are the background and spacer, indicating that the bottom schematic in
\fref[figThreeSiteModel]\letter{A} characterizes the binding of the UP element.
This leads us to the general form of transcriptional regulation by RNAP, shown in \eref[eqGeneExpressionWithUp].
\begin{equation} \label{eqGeneExpressionWithUp}
	\text{GE} = \frac{r_0 + e^{-\beta (E_\text{BG} + E_\text{Spacer} + E_\text{UP})} \left( r_0 e^{-\beta E_{\mhyphen 35}} + r_0 e^{-\beta E_{\mhyphen 10}} + r_{\text{max}} e^{-\beta \left( E_{\mhyphen 35} + E_{\mhyphen 10} + E_\text{int} \right)} \right)}{1 + e^{-\beta (E_\text{BG} + E_\text{Spacer} + E_\text{UP})} \left( e^{-\beta E_{\mhyphen 35}} + e^{-\beta E_{\mhyphen 10}} + e^{-\beta (E_{\mhyphen 35} + E_{\mhyphen 10} + E_\text{int})} \right)}
\end{equation}
\fref[figThreeSiteModel]\letter{B} demonstrates how the expression of all
promoters containing one of the two UP elements combined with each of the eight
background, spacer, -35, and -10 sequences ($2 \times 8^4 = 8{,}192$ promoters)
closely matches the model predictions ($R^2 = 0.88$). Remarkably, since we used
the same free energies and gene expression rates from
\fref[figLogLinearModel]\letter{B}, characterizing these 8,192 promoters only
required two additional parameters (the free energies of the two UP
elements). This result emphasizes how understanding each modular component of
gene expression can enable us to harness the combinatorial complexity of
sequence space.

\subsection*{Sufficiently Strong RNAP-Promoter Binding Energy can Decrease Gene Expression}

Although the 12,288 promoters considered above are well characterized by
\eref[eqGeneExpressionWithUp] on average, the data demonstrate that the full
mechanistic picture is more nuanced. For example, Urtecho \textit{et al.}~found
that gene expression (averaged over all backgrounds and spacers) generally
increases for -35/-10 elements closer to the consensus sequences
\cite{Urtecho2018}. In terms of the gene expression models studied above
(\eref[eqGeneExpressionNoUpApproximate][eqGeneExpressionWithDenominator][eqGeneExpressionNoUp]), promoters with fewer -35/-10 mutations have more negative free energies $E_{\mhyphen 35}$ and $E_{\mhyphen 10}$ leading to larger expression. Yet the strongest promoters with the consensus -35/-10 violated this trend, exhibiting \textit{less} expression than promoters one mutation away. Thus, Urtecho \textit{et al.}~postulated that past a certain point, promoters that bind RNAP too tightly may inhibit transcription initiation and lead to decreased gene expression.

The promoters with a consensus -35/-10 are shown as red points in
\fref[figThreeSiteModel]\letter{B}, and indeed these promoters are all predicted
to bind tightly to RNAP and hence express at the maximum level $r_{\text{max}} =
8.6$, placing them on the right-edge of the data. Yet depending on their UP,
background, and spacer, many of these promoters exhibit significantly less gene
expression then expected. Motivated by this trend, we posit that the state of
transcription initiation can be characterized by a free energy $\Delta
E_{\text{trans}}$ relative to unbound RNAP that competes with the free energy
$\Delta E_{\text{RNAP}}$ between fully bound and unbound RNAP (see Appendix E).

\begin{figure}[t]
	\centering \includegraphics{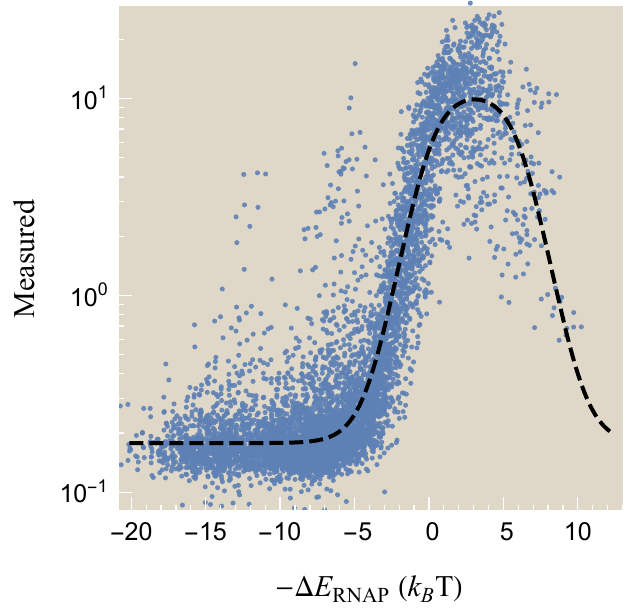}
	
	\caption{\textbf{Gene expression is reduced when RNAP binds a promoter too
			tightly.} Measured gene expression versus the inferred promoter strength
		$\Delta E_{\text{RNAP}}$ relative to the transcription initiation state $\Delta E_{\text{trans}} =
		-6.2\,k_B T$ (stronger promoters on the right). The dashed line shows the
		prediction of the refined energy matrix model.} \label{figStrongPromoters}
\end{figure}

Assuming the rate of transcription initiation is proportional to the relative
Boltzmann weights of these two states, the level of gene expression
$r_{\text{max}}$ in \eref[eqGeneExpressionWithUp] will be modified to
\begin{equation} \label{eqRMaxDependence}
\frac{r_{\text{max}} + r_0 e^{-\beta (\Delta E_{\text{RNAP}} - \Delta E_{\text{trans}})}}{1 + e^{-\beta (\Delta E_{\text{RNAP}} - \Delta E_{\text{trans}})}}.
\end{equation}
As expected, this expression reduces to $r_{\text{max}}$ for promoters that
weakly bind RNAP ($e^{-\beta (\Delta E_{\text{RNAP}} - \Delta E_{\text{trans}})}
\ll 1$) but decreases for strong promoters until it reaches the background level
$r_0$ when the promoter binds so tightly that RNAP is glued in place and unable
to initiate transcription. Upon reanalyzing the gene expression data with the
inferred value $\Delta E_{\text{trans}} = -6.2\,k_B T$ (see Appendix E), we can
plot the measured level of gene expression against the predicted RNAP-promoter
free energy $\Delta E_{\text{RNAP}}$ as shown in \fref[figStrongPromoters]
(stronger promoters to the right). We find that this revised model captures the
downwards trend in gene expression observed for the strongest promoters, most of
which contain a consensus -35/-10.

\subsection*{The Bivalent Binding of RNAP Buffers its Binding Behavior against Promoter Mutations}

In this final section, we investigate how the avidity between the -35 and -10
sites changes the dynamics of RNAP binding. More specifically, we consider the
effective dissociation constant governing RNAP binding when both the -35 and -10
sites are intact and compare it to the case where only one site is capable of
binding. To simplify this discussion, we focus exclusively on the case of RNAP
binding to the -35 and -10 motifs as shown in the rates diagram
\fref[figRNAPoverview]\letter{C}, absorbing the effects of the background,
spacer, and UP elements into these rates.

At equilibrium, there is no flux between the four RNAP states. We define the
effective dissociation constant
\begin{equation} \label{eqEffectiveDissociationRNAP}
K_D^\text{eff} = \frac{K_{\mhyphen 35} K_{\mhyphen 10}}{c_0  + K_{\mhyphen 35} + K_{\mhyphen 10}}
\end{equation}
which represents the concentration of RNAP at which there is a 50\% likelihood
that the promoter is bound (see Appendix F). $K_{j} = \frac{k_{\text{off},
		j}}{k_{\text{on}}}$ stands for the dissociation constant of free RNAP binding to
the site $j$ and $c_0 = \frac{\tilde{k}_{\text{on}}}{k_{\text{on}}} =
[\text{RNAP}] e^{-\beta E_\text{int}}$ represents the increased local
concentration of singly bound RNAP transitioning to the fully bound state (i.e.,
$E_\text{int}$ and $c_0$ are the embodiments of avidity in the language of
statistical mechanics and thermodynamics, respectively). Note that
$K_D^\text{eff}$ is a sigmoidal function of $K_{\mhyphen 10}$ with height
$K_{\mhyphen 35}$ and midpoint at $K_{\mhyphen 10} = c_0 + K_{\mhyphen 35}$.

\fref[figDwellTimeRNAP] demonstrates how the effective RNAP dissociation
constant $K_D^\text{eff}$ changes when mutations to the -10 binding motif alter
its dissociation constant $K_{\mhyphen 10}$. When the -35 sequence is weak
(dashed lines, $k_{\text{off}, \mhyphen 35} \to \infty$), $K_D^\text{eff}
\approx K_{\mhyphen 10}$ signifying that RNAP binding relies solely on the
strength of the -10 site. In the opposite limit where RNAP tightly binds to the -35
sequence (solid lines), the cooperativity $c_0$ and the dissociation constant
$K_{\mhyphen 35}$ shift the curve horizontally and bound the effective
dissociation constant to $K_D^\text{eff} \le K_{\mhyphen 35}$. This upper
bound may buffer promoters against mutations, since achieving a larger effective
dissociation constant would require not only wiping out the -35 site but
in addition mutating the -10 site. Finally, in the case where the cooperativity $c_0$
is large, $K_D^\text{eff} \approx \frac{K_{\mhyphen 10} K_{\mhyphen 35}}{c_0}$
indicating that as soon as one site of the RNAP binds, the other is very likely
to also bind, thereby giving rise to the multiplicative dependence on the two
$K_D$s.

To get a sense for how these numbers translate into physiological RNAP dwell
times on the promoter, we note that the lifetime of bound RNAP is given by $\tau
= \frac{1}{K_D^\text{eff} k_{\text{on}}}$ (see Appendix F). Using
$K_D^\text{eff} \approx 10^{-9}\,\text{M}$ for the strong T7 promoter
\cite{Dayton1984} and assuming a diffusion-limited on-rate $10^8
\frac{1}{\text{M} \cdot \text{s}}$ leads to a dwell time of $10\,\text{s}$,
comparable to the measured dwell time of RNAP-promoter in the closed complex
\cite{Wang2013}. It would be fascinating if recently developed methods that
visualize real-time single-RNAP binding events probed the dwell time of the
promoter constructed by Urtecho \textit{et al.}~to see how well the predictions
of the refined energy matrix model match experimental measurements
\cite{Wang2013}.

\begin{figure}[t]
	\centering \includegraphics{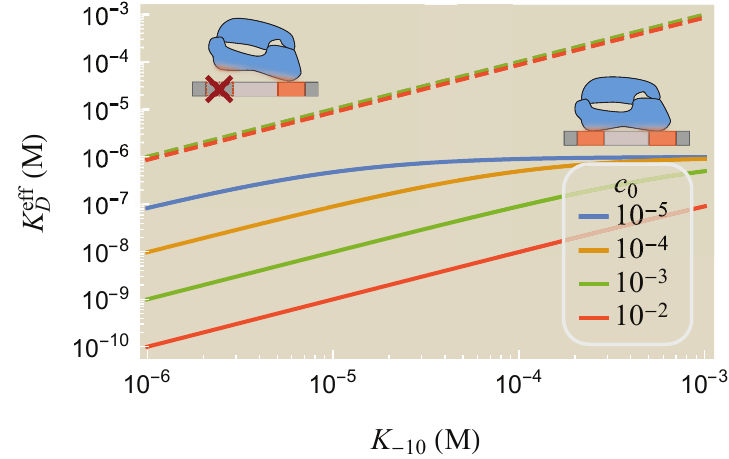}
	
	\caption{\textbf{The dissociation between RNAP and the promoter.}
		RNAP binding to a promoter with a strong (solid lines, $K_{\mhyphen 35} = 1 \mu
		\text{M}$) or weak (dashed, $K_{\mhyphen 35} \to \infty$) -35 sequence. $c_0$
		represents the local concentration of singly bound RNAP.}
	\label{figDwellTimeRNAP}
\end{figure}

\section*{Discussion}

While high-throughput methods have enabled us to measure the gene expression of
tens of thousands of promoters, they nevertheless only scratch the surface of
the full sequence space. A typical promoter composed of 200 bp has $4^{200}$
variants (more than the number of atoms in the universe). Nevertheless, by
understanding the principles governing transcriptional regulation, we can begin
to cut away at this daunting complexity to design better promoters.

In this work, we analyzed a recent experiment by Urtecho \textit{et
	al.}~measuring gene expression of over 10,000 promoters in \textit{E.~coli}
using the $\sigma^{70}$ RNAP holoenzyme \cite{Urtecho2018}. These sequences
comprised all combinations of a small set of promoter elements, namely, eight
-10s, eight -35s, eight spacers, eight backgrounds, and three UPs depicted in
\fref[figRNAPoverview]\letter{A}, providing an opportunity to deepen our
understanding of how these elements interact and to compare different
quantitative models of gene expression.

We first analyzed this data using classic energy matrix models which posit that
each promoter element contributes independently to the RNAP-promoter binding
energy. As emphasized by Urtecho \textit{et al.}~and other groups, such energy
matrices poorly characterize gene expression
(\fref[figLogLinearModel]\letter{A}, $R^2 = 0.57$) and offer testable
predictions that do not match the data (Appendix C), mandating the need for
other approaches \cite{Urtecho2018, Forcier2018}.

To meet this challenge, we first determined which promoter elements contribute
independently to RNAP binding (Appendix D). This process, which was done without
recourse to fitting, demonstrated that the -35 and -10 elements bind in a
concerted manner that we postulated is caused by avidity. In this context,
avidity implies that when RNAP is singly bound to either the -35 or -10 sites,
it is much more likely (compared to unbound RNAP) to bind to the other site,
similar to the boost in binding seen in bivalent antibodies \cite{Klein2010} or
multivalent systems \cite{Banjade2014, Huang2016, Yan2018}. Surprisingly, we
found that outside the -35/-10 pair, the other components of the promoter
contributed independently to RNAP binding.

Using these findings, we developed a refined energy matrix model of gene
expression (\eref[eqGeneExpressionWithUp]) that incorporates the avidity of
between the -35/-10 sites as well as the independence of the
UP/spacer/background interactions. This model was able to characterize the 4,096
promoters with no UP element (\fref[figLogLinearModel]\letter{B}, $R^2 = 0.91$)
and the 8,192 promoters containing an UP element
(\fref[figThreeSiteModel]\letter{B}, $R^2 = 0.88$). These results surpass those
of the traditional energy matrix model (\fref[figLogLinearModel]\letter{A}, $R^2
= 0.57$), only requiring two additional parameters that could be experimentally
determined (e.g., the interaction energy $E_\text{int}$ arising from the -35/-10
avidity and the level of gene expression $r_0$ of a promoter with a scrambled
-10 motif, a scrambled -35 motif, or with both motifs scrambled).

These promising findings suggest that determining which components bind
independently is crucial to properly characterize multivalent systems. It would
be fascinating to extend this study to RNAP with other $\sigma$ factors
\cite{Feklistov2014} as well as to RNAP mutants with no $\alpha$CTD or that do
not bind at the -35 site \cite{Kumar1993, Minakhin2003}. Our model would predict
that polymerases in this last category with at most one strong binding site
should conform to a traditional energy matrix approach.

Quantitative frameworks such as the refined energy matrix model explored here
can deepen our understanding of the underlying mechanisms governing a system's
behavior. For example, while searching for systematic discrepancies between our
model prediction and the gene expression measurements, we found that promoters
predicted to have the strongest RNAP affinity did not exhibit the largest levels
of gene expression (thus violating a core assumption of nearly all models of
gene expression that we know of). This led us to posit a characteristic energy
for transcription initiation that reduces the expression of overly strong
promoters (\fref[figStrongPromoters]). In addition, we explored how having
separate binding sites at the -35 and -10 elements buffers RNAP kinetics against
mutations; for example, no single mutation can completely eliminate gene
expression of a strong promoter with the consensus -35 and -10 sequence, since
at least one mutation in both the -35 and -10 motifs would be needed
(\fref[figDwellTimeRNAP]).

Finally, we end by zooming out from the particular context of transcription
regulation and note that multivalent interactions are prevalent in all fields of
biology \cite{Gao2018}, and our work suggests that differentiating between
independent and dependent interactions may be key to not only characterizing
overall binding affinities but to also understand the dynamics of a system
\cite{Stone2011}. Such formulations may be essential when dissecting the much
more complicated interactions in eukaryotic transcription where large complexes
bind at multiple DNA loci \cite{Goardon2006, Levine2014} and more broadly in
multivalent scaffolds and materials \cite{Varner2015, Yan2018}.

\section*{Methods}

We trained both the standard and refined energy matrix models on 75\%
of the data and characterized the predictive power on the remaining 25\%,
repeating the procedure 10 times. The coefficient of determination $R^2$ was
calculated for $y_{\text{data}} = \log_{10}(\text{gene expression})$ to prevent
the largest gene expression values from dominating the result. The supplementary
Mathematica notebook contains the data analyzed in this work and can recreate
all plots.

\section*{Acknowledgements}

We thank Suzy Beeler, Vahe Galstyan, Peng (Brian) He, and Zofii
Kaczmarek for helpful discussions. This work was supported by the Rosen Center
at Caltech and the National Institutes of Health through 1R35 GM118043-01
(MIRA).

\printbibliography

\end{document}